# Suspending Effect on Low-Frequency Charge Noise in Graphene Quantum Dot

Xiang-Xiang Song,[1,2] Hai-Ou Li,[1,2] Jie You,[1,2] Tian-Yi Han,[1,2] Gang Cao,[1,2] Tao Tu,[1,2] Ming Xiao,[1,2] Guang-Can Guo,[1,2] Hong-Wen Jiang,[3] and Guo-Ping Guo[1,2]

[1] *Key Laboratory of Quantum Information, CAS, University of Science and Technology of China, Hefei, Anhui 230026, China*
[2] *Synergetic Innovation Center of Quantum Information & Quantum Physics, University of Science and Technology of China, Hefei, Anhui 230026, China*
[3] *Department of Physics and Astronomy, University of California at Los Angeles, CA 90095, USA*

Charge noise is critical in the performance of gate-controlled quantum dots (QDs). Such information is not yet available for QDs made out of the new material graphene, where both substrate and edge states are known to have important effects. Here we show the 1/f noise for a microscopic graphene QD is substantially larger than that for a macroscopic graphene field-effect transistor (FET), increasing linearly with temperature. To understand its origin, we suspended the graphene QD above the substrate. In contrast to large area graphene FETs, we find that a suspended graphene QD has an almost-identical noise level as an unsuspended one. Tracking noise levels around the Coulomb blockade peak as a function of gate voltage yields potential fluctuations of order 1 μeV, almost one order larger than in GaAs/GaAlAs QDs. Edge states and surface impurities rather than substrate-induced disorders, appear to dominate the 1/f noise, thus affecting the coherency of graphene nano-devices.

*Correspondence and requests for materials should be addressed to G.P.G. (gpguo@ustc.edu.cn)

Low frequency 1/f charge noise plays a significant role in modern electronics.[1] Although the origin of the 1/f noise is not well known, it is believed that the randomly changing charge distribution of electron traps in the device results in a 1/f dependence.[1,2] For quantum devices, the charge noise is generally regarded as the major de-coherence source of charge-state-encoded qubits.[3–6,36] Because of its unique properties, such as the absence of residual nuclear spin and weak spin-orbital coupling,[7–9] graphene has attracted much attention for its promising variety of electronics applications. Much research has focused on charge-noise measurements of graphene devices.[10–12,42] The edge states and disorders induced during fabrication and the device substrate can greatly affect the properties of graphene devices [13–16]. For example, suspended graphene flakes can yield a huge increase in low-temperature mobility approaching 200,000 cm$^2$ V$^{-1}$ S$^{-1}$ for carrier densities below $5 \times 10^9$ cm$^{-2}$.[17,18] The charge noise of graphene field effect transistors (GFETs) can be suppressed by one order of magnitude when suspended from the substrate.[11]

However, most of the charge noise experiments had focused on macroscopic graphene devices such as micrometer-sized GFETs. There has been no report on the charge noise for graphene nano-devices where the carrier channel size is of 10-nm order although researchers have fabricated various graphene nanostructures including single graphene quantum dots (GQDs),[19–25] double GQD, both in series and in parallel,[26–30] and the hybrid system of GQD and superconductor cavity.[32] Recently, the relaxation time $T_1$ (~100 ns) and the dephasing time $T_2$ (~1 ns) have both been measured for the charge states in double GQDs.[31,32] The coherence times of these charge state are on the same order of magnitude as in the traditional semiconductor double quantum dots. To improve the electronic performance and increase the quantum coherence in graphene nanostructures, knowing the exact level of the charge noise and how experimental conditions such as substrate and electron temperature affect this noise is valuable. In this letter, we used the wet-etching method[33,34] to fabricate suspended graphene nanoribbons. These suspended nanoribbon devices behave similarly to unsuspended ones and can also be tuned to the Coulomb blockade region to form GQDs. By measuring the 1/f noise along Coulomb peaks, we can obtain the charge noise, corresponding to potential energy fluctuations, which increase linearly with temperature. Different from GFETs, there is no observable change in charge noise by suspending GQDs from their substrate. Both suspended and unsuspended GQD devices have similar charge noise of the order of 1 μeV , which is one order larger than that in GaAs/AlGaAs QD[35] and two orders larger than that in GFETs[11,42]. We present a simplified model explaining how edge states rather than the substrate act as more important sources of charge noise in GQD devices, inferring that the elimination of edge states should be a key task for the future when exploiting graphene nano-devices.

# Results

## Suspended graphene quantum dots

Fig. 1a shows a scanning electron microscope (SEM) image of the graphene nano-device. The graphene nanoribbon is about 200 nm long and has a width less than 100 nm. Two source-drain contacts are 800 nm apart. The side gate is 200 nm away from the graphene and is used in tuning the electronic potential in the nanoribbon.

After dipping the sample in the BOE for 40 seconds, $SiO_2$ was etched away to a depth of about 50 nm. We estimate our graphene nanoribbon was suspended about 50 nm above the substrate. By tilting the sample holder of the SEM to a specified angle, we can observe the nano-device from the side to check whether the graphene nanoribbon was suspended. As shown in Fig. 1b, the ribbon appears flat above the substrate. This feature is very different from former situations, where we clearly observed graphene flakes bent on the substrate in contact with the substrate (see Supporting Information Fig. S1 and S2). The experiment was performed in a He3 refrigerator at a base temperature of 240 mK.

We used the standard lock-in method to probe the electronic signals to make the suspended nanoribbon work in the Coulomb blockade region. We applied a dc voltage to the back gate to tune the Fermi energy of the graphene. The V-shaped current-voltage relation was observed, and a transport gap was found in the range 0.5–6.0 V (Fig. 2a). We then changed the voltage between the source and drain, and measured the tunneling current at different back gate voltages within the transport gap, often known as the Coulomb-diamonds measurement. We found the largest charging energy exceeding 5 meV (Fig. 2b), indicating the suspended graphene nanoribbon worked as a single dot. The side gate was grounded in this measurement.

## Measurement of the charge noise

We measured the current fluctuation using a spectrum analyzer (SR785). A schematic of the circuit used is shown in Fig. 1a. As we swept the back gate voltage, we obtained different spectra. Fig. 3b shows three different noise spectra, labeled A, B, C in Fig. 3a, measured from different parts of the Coulomb peak. The spectrum at B shows clear 1/f dependence up to 100 Hz. The spectrum at A shows lower noise but also with a 1/f dependence. The spectrum at C was measured in the Coulomb blockade region, where only a very small tunneling current can be probed. This spectrum represents the noise of our measurement system, including amplifiers and contacts. All the data points near 50 Hz were removed as they were induced by electricity from the mains.

We calculated the magnitude of the current fluctuation $\Delta I$ by integrating the spectrum from 1 to 9 Hz;

$$\Delta I = \sqrt{\int_1^9 [S_I^2(f) - S_{CB}^2(f)] df}. \qquad (1)$$

In equation, $S_I(f)$ is the noise spectrum measured. $S_{CB}(f)$ is the background noise induced by the measurement system. Here we used the spectrum at C as $S_{CB}$.

Next, we applied a dc voltage $V_{SD}$=200 μV to the source and measured the current from the drain. Changing the back gate voltage from 4.283 to 4.297 V, we obtained a Coulomb peak. Fig. 4a shows the current-gate voltage relation. We calculated the derivative of the current, |dI/dV|, which is plotted in Fig. 4b. Then, we measured the spectrum $S_I$ at different back-gate voltages along the Coulomb peak and calculated the integral in Equation (1). The integrated current fluctuation ΔI is presented in Fig. 4c. We note that ΔI has two peaks almost at the same gate voltage where the derivative of the current has maximum value. This feature has been observed before in Ref. [35], indicating that the fluctuation of the potential is dominant here. The nonzero current noise in the region between the two peaks corresponds to the fluctuation of the tunneling rate, ΔΓ. Here, we found ΔΓ was much larger than that in GaAs devices.[35] Finally, using the relation $ΔI=α^{-1}|dI/dV|Δε$ given in Ref. [35], we subtracted the dependence on the derivative of the current and yielded the fluctuation of the potential in terms of energy. Here, α is a conversion factor from the back gate voltage to the potential energy, known as the lever arm. In this way, we can obtain a parameter Δε in the energy scale, which is convenient for comparison. Note Δε is also independent of current, which means the magnitude of potential fluctuation should vary within a particular range when back gate voltage changes over a relatively long range. The relation between ΔI and Δε can be understood in terms of the normalization of the noise measured along the Coulomb peak. The parameter Δε can be regarded as a reasonable parameter for describing the overall noise level of the nano-devices.

The potential fluctuation Δε, calculated from the peak in Fig. 4a, is shown in Fig. 5a, using blue squares (labeled as Device#1). Note that the value of Δε can be large if the value of |dI/dV| in the denominator is small. This inaccuracy is just a result of the calculation. Hence, we only used data points where the derivative of the current is not small, which means the region between the two dotted lines shown in Fig. 4a–c. The data points of the pinnacle were also no considered for the same reason. After removing these data points, the magnitude of the fluctuation was found in the range 0.75−1.5 μeV. From the Coulomb diamond measurement, we estimated that the lever arm of the back gate is at about 0.07 eV/V for suspended graphene nano-devices.

We also measured the current noise of regular (unsuspended) graphene nanoribbon devices for comparison. The regular graphene device was fabricated in a similar way as described above. The only difference was that the regular samples were not dipped in BOE after making electrodes. Similar noise measurements of regular devices were performed under the same experiment conditions. Fig. 4d shows a Coulomb peak measured in a regular graphene nanoribbon device. We also obtained |dI/dV| and ΔI for the peak (see Fig. 4e and 4f). We estimated that the lever arm of our unsuspended nano-device is at about 0.12 eV/V, almost twice as much as that of the suspended device.

Similarly, the potential fluctuation was calculated in the range 0.5−1.5 μeV (open blue triangles (labeled as Device#5) in Fig. 5a).

To investigate the influence of the temperature on the potential fluctuation, we measured the same Coulomb peak at different temperatures, from 240 mK to 1 K. From Fig. 5b, the Coulomb peak becomes broader and higher. We also measured Δε at different temperature. With increasing temperature, the potential fluctuation increases almost linearly (Fig. 5c), showing the simple 1/f noise model is still valid for our nano-device at low temperature.

We measured different Coulomb peaks from several devices, both suspended and unsuspended. For Fig. 5a, all the results we obtained were in the range 0.5−2.8 μeV. The difference in the potential fluctuation between suspended and regular devices of the same topological design pattern was within one order of magnitude. From these experiments, we found that for the microscopic GQD fabricated by etching, the substrate has no significant effect on low-frequency noise. This result is different from the report for the GFET, where it is shown that the removal of the substrate can decrease the low-frequency noise by one order of magnitude.[11] We also compared the potential fluctuation of graphene nano-devices to that from GaAs/AlGaAs heterostructures. The Δε of GaAs quantum dots ranges from 0.07 to 0.16 μeV (black stars in Fig. 5a),[35] which is clearly one order of magnitude lower than our results. As $T_2^* \propto 1/\sqrt{\Delta\varepsilon}$,[5,6] this results in a shorter $T_2^*$ in graphene nano-devices. Notably, we also used the method described in Ref. [11] and Ref. [42] to compare the noise of our microscopic GQD with the macroscopic GFET. We found the noise of a GQD is one to two orders larger than that of a GFET (see Supporting Information Fig. S3).

## Discussion

It is well known that the edge states can greatly affect the properties of graphene devices. The edge states change the electron distribution, resulting in the formation of puddles in graphene nano-devices.[37–39] Tunneling through these puddles influences the noise spectra as well. Here, we present a simple model to explain our results. Assuming the carrier density is N, the edge states density is $n_E$, the density of disorders in the substrate is $n_S$. The effect of edge states and substrate on the low frequency noise of the device is proportional to $\frac{1}{(\overline{r_E})^2}\frac{n_E}{N}$ and $\frac{1}{(\overline{r_S})^2}\frac{n_S}{N}$, respectively. Here, $\overline{r_E}$ ($\overline{r_S}$) is mean effective interaction distance between carriers and edge states (disorders in the substrate). Obviously, the edge states in a microscopic GQD are much closer to a carrier channel, which means a smaller interaction distance $\bar{r}_E$, results in a greater interaction on the carriers than in a macroscopic GFET because of its small size. Here, we simply assume $\bar{r}_S$ of a GFET and a GQD are the same. The difference in suspended and regular devices is the absence (presence) of $n_S$. In the GFET experiment, the removal of the substrate decreases the noise by one order of magnitude.[11] However, for the GQD, the influence of the edge states increases rapidly because of the smaller $\bar{r}_E$. Moreover, the carrier is moved through GQD one by one to show Coulomb peaks, which means a lower carrier density N in GQD, resulting in larger noise as well. As the influence caused by edge states increases, the effect from the substrate is no longer dominant here, showing almost no difference between suspended and regular nano-devices. In traditional semiconductor GaAs gate-defined QDs, the fact that the low-frequency noise is almost one order of magnitude lower than in GQDs can also be understood from the absence of edge states. Note that surface impurities may also contribute to scattering and to 1/f noise. Since fabrication induced residue may contaminate the devices, resulting in defects on the surface of graphene flakes. Tunneling through these defects may also be a noise source of graphene QDs. We cannot exclude the influence of these surface impurities in our experiment.

Although this simple model can qualitatively explain the present experiment, more studies are still needed to investigate how the edge states or surface impurities affect the 1/f noise, and how to decrease the noise level in the graphene devices.

In summary, we have fabricated both suspended and unsuspended graphene nano-devices and measured their 1/f noise along the Coulomb peaks to obtain the charge noise level. Suspending the QD from the substrate had no observable effect on the charge noise of a GQD. The edge states closely surrounding these nano-devices are argued to increase the charge noise to 1 μeV and become the dominant charge noise source instead of that from the substrate. More studies are needed to improve the electronic performance and increase the quantum coherence of future graphene nano-devices.

## Methods

Graphene flakes were produced by mechanical cleaving of bulk graphite and deposited on a highly doped silicon substrate covered by 100 nm of silicon dioxide. The doped silicon substrate worked as a back gate. Graphene flakes were found using an optical microscope, and few-layer flakes were selected using Raman spectroscopy.[40,41] After depositing the graphene flakes on the substrate, we used polymethyl methacrylate (PMMA) in the standard electron beam lithography technique to form the designed pattern. The unprotected parts of the graphene were removed by inductive coupling plasma (ICP). A second electron beam lithography process followed by e-beam evaporation of Ti/Au was used to make both the source-drain contacts and the side gate. Finally, we dipped the sample in buffered oxide etch (BOE) for 40 s to etch part of the SiO$_2$ layer away. To avoid graphene from being wrinkled, we used a critical-point dryer to dry the sample instead of blowing with N$_2$.

## Acknowledgments

This work was supported by the National Fundamental Research Program (Grant No. 2011CBA00200), and National Natural Science Foundation (Grant Nos. 11222438, 11174267, 11304301, 61306150, 11274294, and 91121014) and the Chinese Academy of Sciences.

## Author contributions

X.X.S. fabricated the samples. X.X.S., H.O.L., J.Y., T.Y.H., G.C. and G.P.G. performed the measurements. X.X.S., M.X., T.T. and G.C.G. provided theoretical support. Data are analyzed by X.X.S., H.O.L. and J.Y. The manuscript is prepared by X.X.S., H.W.J. and G.P.G., G.P.G. supervised the project. All authors contributed in discussing the results and commented on the manuscript.

## Additional information

Supplementary information, which contains further information on the devices properties, noise analysis using the method described in Ref. [11], is available at xxxxxxxxx.
Competing financial interests: The authors declare no competing financial interests.

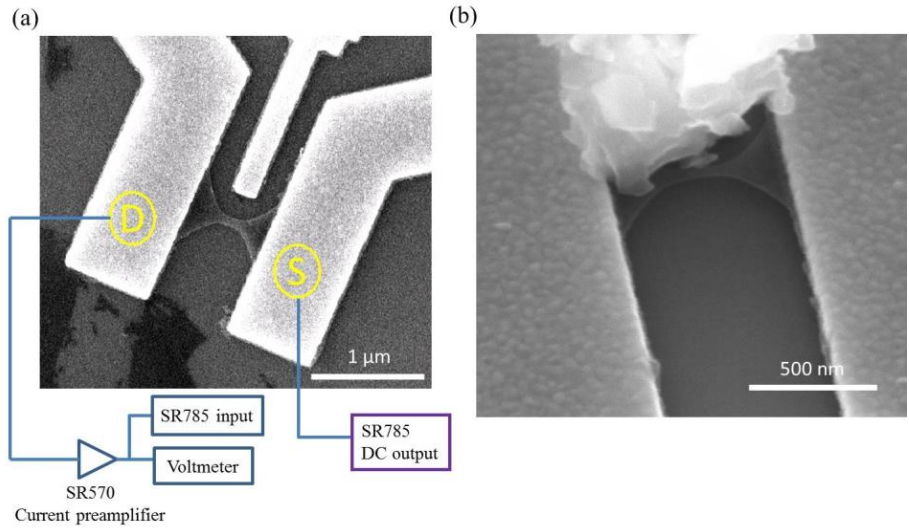

**Figure 1. Device characterizaion.** (a) SEM image of a suspended graphene nanoribbon device and schematic of the circuit used in noise measurement. The white bar has a length of 1 μm. (b) Zoom in of a similarly-fabricated sample, indicating that the graphene is suspended. (The white cluster above is a particle of dust fallen on the sample when taking the sample off the chip carrier after measurement.) The white bar here indicates a length of 500 nm.

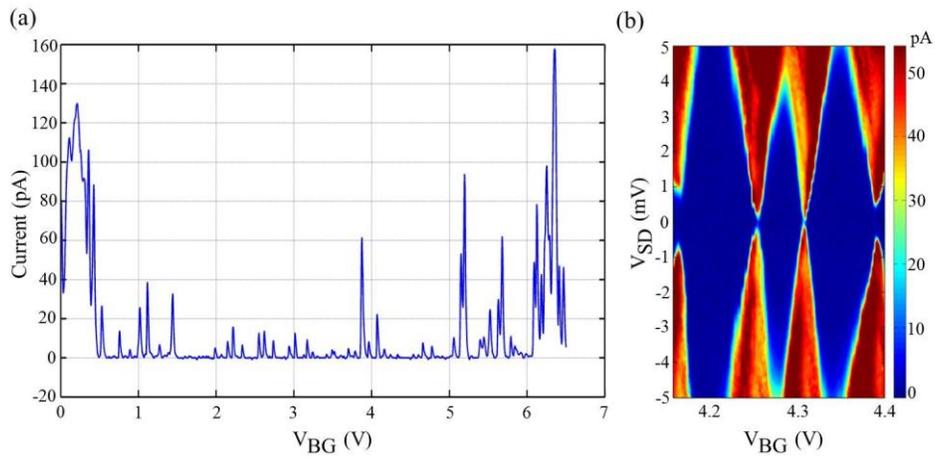

**Figure 2. Transport measurements of the device.** (a) Source-drain tunneling current flows through the suspended graphene nanoribbon device as a function of back gate voltage $V_{BG}$. The back gate voltage ranges from 0 to 6.5 V. A transport gap is found from 0.5 to 6 V. (b) Coulomb diamonds measured within the transport gap. The largest charging energy exceeds 5 meV. The side gate is grounded in this measurement.

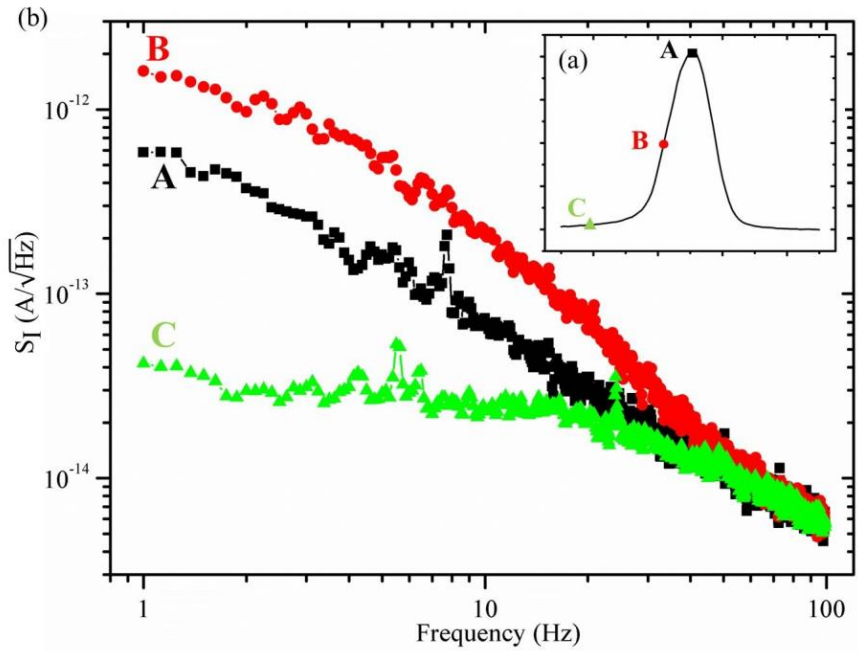

**Figure 3. Noise spectra measured from Coulomb peaks.** (a) A typical tunneling current peak, known as the Coulomb peak, measured when the back gate voltage $V_{BG}$ is swept. (b) Noise spectra measured from three different regions of the Coulomb peak, labeled A, B, C in (a). The figure is plotted in the log-log scale.

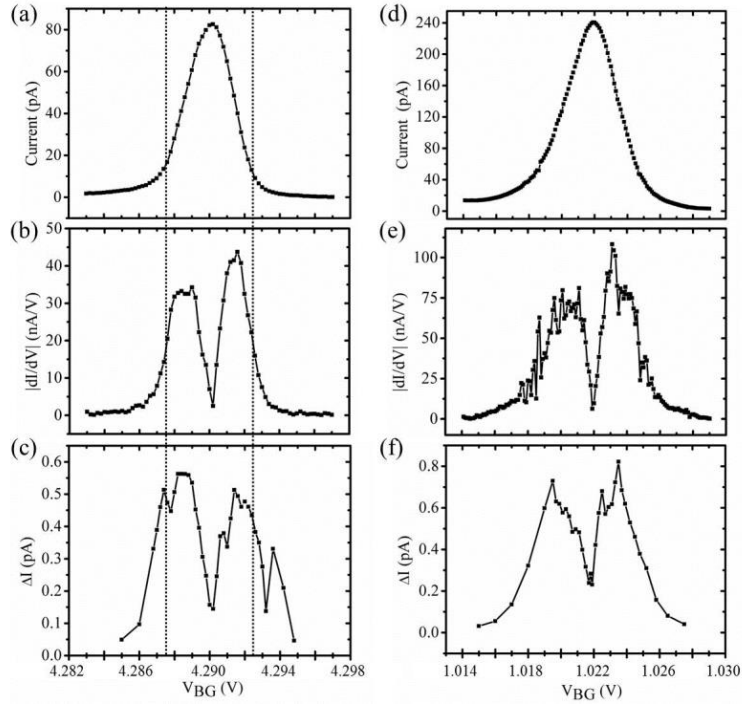

**Figure 4. Normalization of the noise.** (a) A Coulomb peak obtained from a suspended graphene nanoribbon at $V_{SD}=200\,\mu V$ when back gate voltage ranges from $V_{BG}=4.283\,V$ to $V_{BG}=4.297\,V$. (b) The derivative of the current, $|dI/dV|$, of the peak shown in (a). (c) The current fluctuation $\Delta I$ of the peak shown in (a), as a function of back gate voltage $V_{BG}$. To avoid inaccuracy, we only used the data points between two dotted lines in (a), (b), and (c) for calculating the potential fluctuation $\Delta\varepsilon$. (d) A Coulomb peak obtained from an unsuspended graphene nanoribbon. Back gate voltage ranges from $V_{BG}=1.014\,V$ to $V_{BG}=1.029\,V$ where $V_{SD}=350\,\mu V$. (e) The derivative of the current of the peak shown in (d). (f) The current fluctuation $\Delta I$ of the peak shown in (d) as a function of back gate voltage $V_{BG}$.

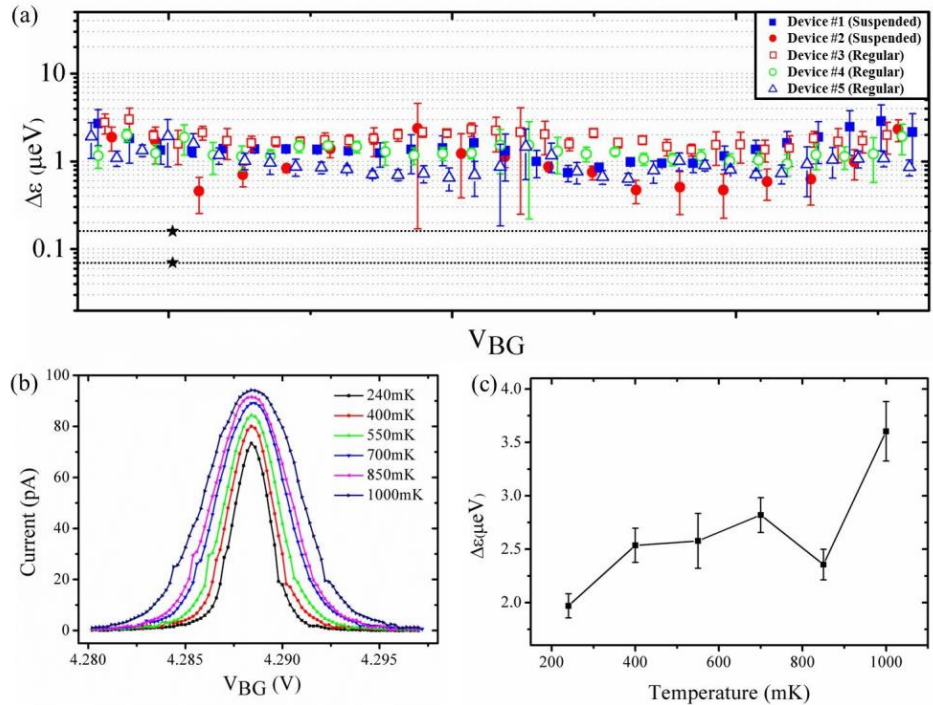

**Figure 5. Noise level of suspended and regular graphene nano-devices and its temperature dependence.** (a) The potential fluctuation Δε, as a function of $V_{BG}$, obtained from five different graphene nano-devices, including both suspended and regular. All curves have been resized to the same x-axis scale for greater visibility. The black stars correspond to the result of GaAs devices described in Ref. [35]. (b) A typical Coulomb peak measured at different temperatures. (c) The potential fluctuation Δε as a function of temperature.

# Supporting Information for

## Suspending Effect on Low-Frequency Charge Noise in Graphene Quantum Dot

Xiang-Xiang Song, [1, 2] Hai-Ou Li, [1, 2] Jie You, [1, 2] Tian-Yi Han, [1, 2] Gang Cao, [1, 2]
Tao Tu, [1, 2] Ming Xiao, [1, 2] Guang-Can Guo, [1, 2] Hong-Wen Jiang, [3] and Guo-Ping Guo[1, 2]

[1] *Key Laboratory of Quantum Information, CAS, University of Science and Technology of China, Hefei, Anhui 230026, China*

[2] *Synergetic Innovation Center of Quantum Information & Quantum Physics, University of Science and Technology of China, Hefei, Anhui 230026, China*

[3] *Department of Physics and Astronomy, University of California at Los Angeles, CA 90095, USA*

## 1. Suspension of the device

Figure S1 shows a SEM image of a graphene nanoribbon, which is in contact with the substrate. (This device is not used in the experiment.) Figure S2(a) shows a SEM image of a suspended graphene nanoribbon device after transport measurement. Clearly, the nanoribbon is not pulled down to the substrate during the transport measurement, which is different from the situation in Figure S1. Even though a dc voltage of up to 7 V was applied to the back gate, the graphene nanoribbon still remains suspended. Figure S2(b) shows a SEM image of the same device in Figure S2(a) after AFM measurement. The graphene nanoribbon is tapped down, in contact with the substrate, which is similar with Figure S1.

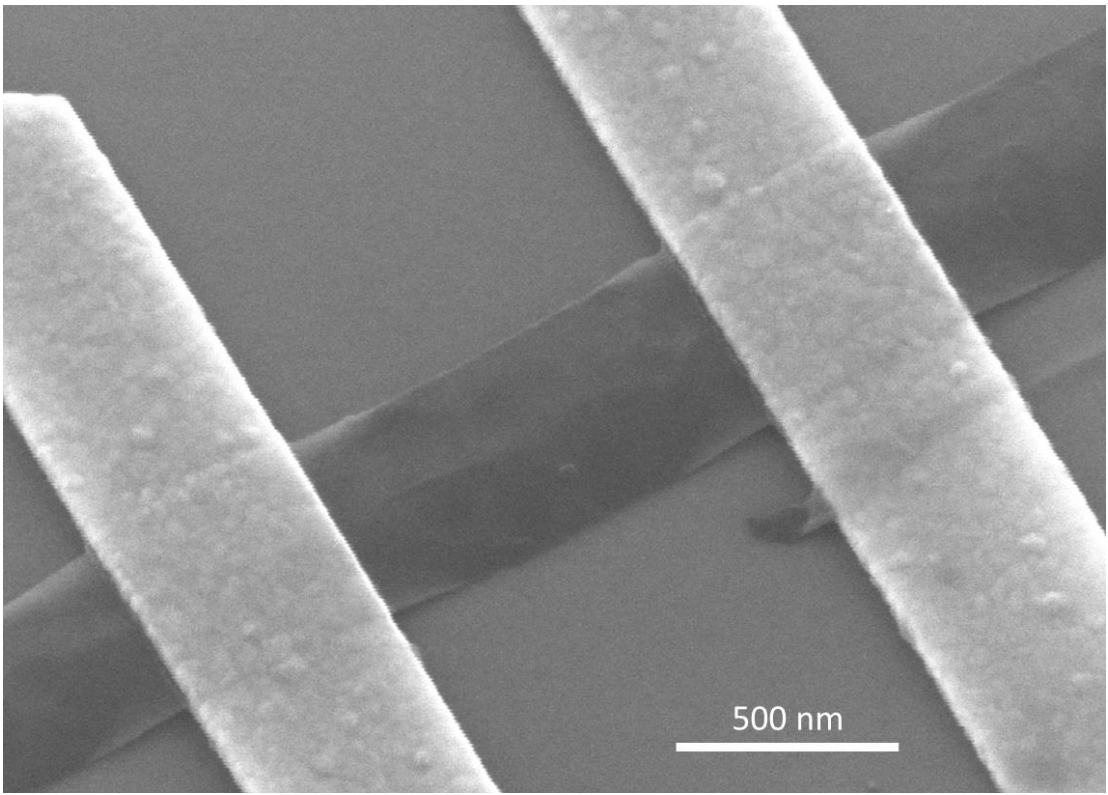

Figure S1. SEM image of a graphene nanoribbon in contact with the substrate surface.

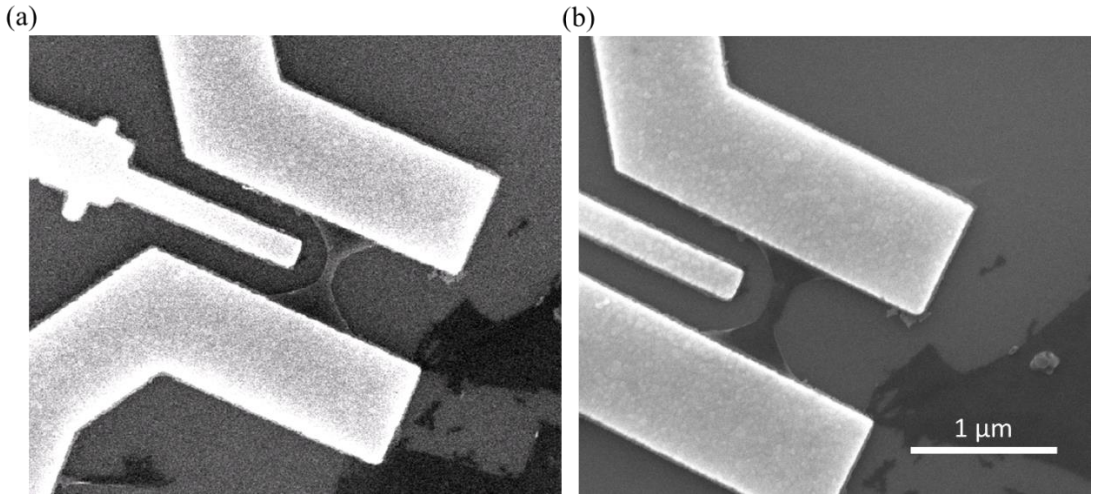

Figure S2. (a) SEM image of a suspended graphene nanoribbon device after transport measurement. (b) SEM image of the same device after AFM measurement.

## 2. Lever arm

We measure the noise level of five graphene nano-devices, both suspended and regular (unsuspended). As suspend the device above the substrate, the electrical environment is changed, resulting in the difference between the lever arms of the device.

We obtain the lever arm from standard Coulomb diamonds measurement.[S1] As shown in Table S1, we find regular devices have a lever arm almost twice as much as those of suspended devices.

| Device Number | Device properties | Lever arm (eV/V) |
| --- | --- | --- |
| #1 | Suspended | 0.07 |
| #2 | Suspended | 0.08 |
| #3 | Regular | 0.12 |
| #4 | Regular | 0.11 |
| #5 | Regular | 0.12 |

Table S1. Lever arms of five different nano-devices labeled in Figure 5.

## 3. Noise spectra analysis using another method

Figure S3 shows two different noise spectra, calculated using the method described in Ref. [S2]. The spectra (red and black) were measured at different regions of the Coulomb peak, labeled A and B in the inset (similar as in Figure 3). The figure is plotted in log-log scale. All the data points near 50 Hz were removed as they were induced by electricity from the mains. The blue and green dashed lines are the noise spectra obtained from regular and suspended graphene FETs respectively (see Ref. [S2]), showing a difference of an order of magnitude. However, in our experiment, no difference was observed between regular and suspended graphene QDs using this method.

To compare our results to graphene FETs, we focus our attention on spectrum A, since electrons tunnel through the graphene QD at A. Clearly, the noise at A is one (two) order(s) of magnitude larger than the result obtained from regular (suspended) GFETs, respectively.

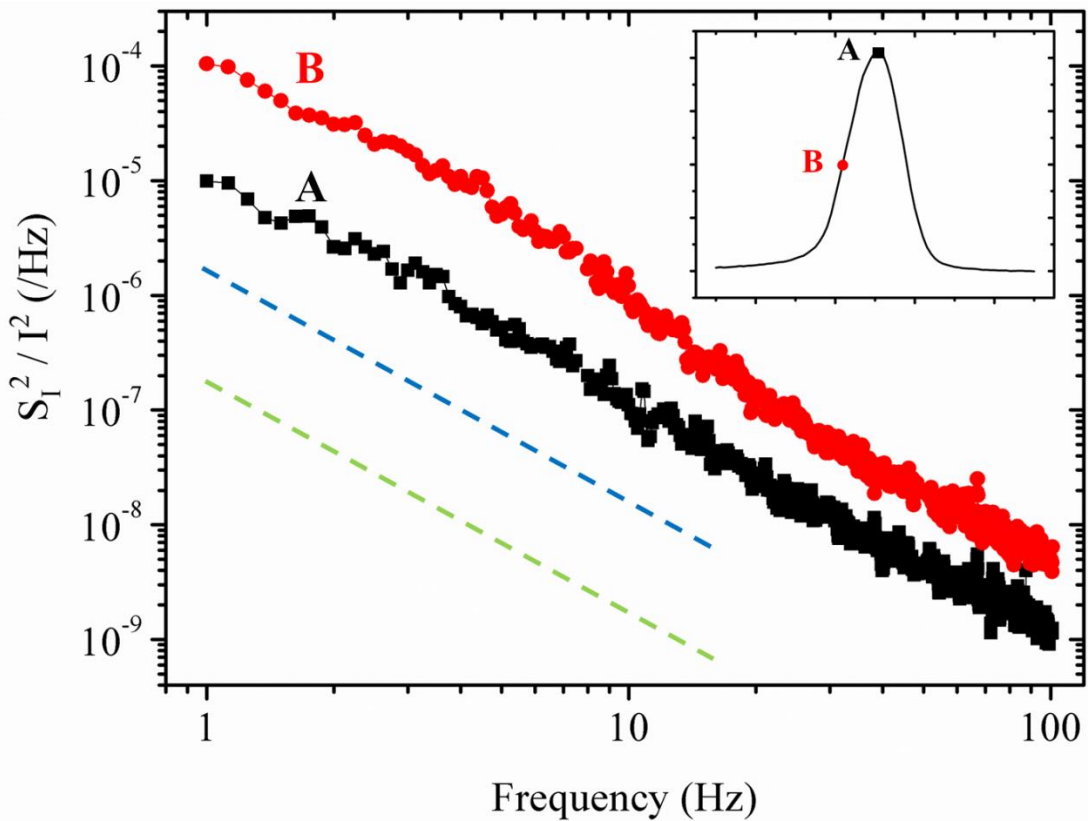

Figure S3. Two different noise spectra measured at different regions of the Coulomb peak, labeled A and B in the inset. The figure is plotted in log-log scale. The blue and green dashed lines are the noise spectra obtained from regular and suspended graphene FETs, respectively (see Ref. [S2]).

We also compared our results to which described in Ref. [S3]. Fitting the spectra with the formula $S_I^2/I^2=A/f$, we find that the noise power A of our graphene nano-devices (both suspended and unsuspended) is of the order of $10^{-6}$ to $10^{-5}$, which is about one (two) order(s) of magnitude larger. Furthermore, we obtained our area-scaled noise amplitude (device area ($\mu m^2$)×noise power A), as $10^{-2} \times 10^{-5}$ ($10^{-6}$)=$10^{-7}$ ($10^{-8}$). Since we only consider the area of the nanoribbon, the effective area should also include the area of the connection part between the nanoribbon and source-drain contacts, results in larger area-scaled noise amplitude. Compared to the Figure 3 in Ref. [S3], our result is one (two) order(s) of magnitude larger. This result is also consisted with the comparison in the Figure S3, indicating some new sources of noise, such as edge states and surface impurities, influence the performance of the graphene QDs.